\documentclass[multi]{cambridge6A}
\usepackage{graphicx}
\usepackage{amsmath,amssymb}
\begin{document}
\newcommand{\be}{\begin{equation}}
\newcommand{\ee}{\end{equation}}
\newcommand{\bea}{\begin{eqnarray}}
\newcommand{\eea}{\end{eqnarray}}

\author[Ruth Gregory]{Ruth Gregory}

\chapter{The Gregory-Laflamme instability}

\contributor{Ruth Gregory \affiliation{Durham University}} 

 \index{Gregory-Laflamme instability|see {instability, black string}}

In this chapter\footnote{Chapter 
of the book {\it Black Holes in Higher Dimensions} to be 
published by Cambridge University Press (editor: G. Horowitz).} 
we introduce the notion of higher dimensional gravity
in the context of one extra spatial dimension. We focus on the {\it black 
string}: a simple extension of the Schwarzschild solution into five 
dimensions. Here, we will show that this solution is unstable to
long wavelength perturbations, and discuss its implications and
extensions.

\section{Overview}

Kaluza \cite{Kaluza2} and Klein \cite{Klein2}, very shortly after
Einstein's general relativity had been verified by Eddington's 1919 expedition,
suggested that adding an extra dimension to our space could have
an amazing consequence: `Gravity' in five dimensions with the
extra dimension stabilized has the appearance of Einstein-Maxwell
theory in a four dimensional slice. Kaluza-Klein theory, as
it is now known, is a construction adding extra dimensions to
space which are much smaller than scales we can directly physically
probe, and thus contribute only a few long range, or massless, additional
forces to nature. Kaluza-Klein theory is reviewed in full in Chapter 4, but
for the purposes of this chapter, we will only require some very 
basic intuition. We will consider only solutions to vacuum gravity
in five dimensions, and focus on a particularly simple solution: the black
string. We will describe the solution, discuss its properties, then
demonstrate explicitly that it is unstable to linear perturbations.
We conclude with a brief discussion of the more general situation.

\section{Black holes in higher dimensions}

The Schwarzschild solution in four dimensions is found by solving 
the vacuum Einstein equations
\be
R_{\mu\nu} = 0\;,
\ee
subject to a physically motivated spherical symmetry restriction. 
It is known that in four dimensions, the only possible static black 
hole solution must be spherically symmetric -- but what happens 
if we live in four, or more, {\it spatial}
dimensions? In particular, what happens if that final spatial
dimension has finite extent and is very small?

In five dimensions, we obviously have to solve the same
equation\footnote{We reserve the labels $\mu$, $\nu\dots$
for purely four dimensional indices, and use $a$, $b\dots$ for
the full range of dimensions.}
\be
R_{ab} = 0 \; .
\ee
However, we now have to choose an appropriate symmetry
for the spacetime metric. An obvious generalisation of the
Schwarzschild solution is to take
a hyperspherically symmetric solution:
\be
ds^2 = -V_5(r) dt^2 + V_5(r)^{-1} dr^2 + r^2 d\Omega_3^2\;,
\label{5DSCH}
\ee
where $V_5$ is an appropriate generalisation of the four dimensional 
Schwarschild potential. This is indeed the case (see Chapter 5 or
 \cite{MP2} for a discussion of general solutions), 
if the extra dimension is infinite, when $V_5$ is given explicitly by
\be
V_5(r) = 1 - \frac{r_5^2}{r^2}\;,
\label{5Dpot}
\ee
where $r_5$ is the horizon radius, and is related to the mass
of the black hole via \cite{MP2},
\be
r_5^2 = \frac{8 G_5 M_5}{3 \pi}\;.
\label{M5}
\ee
However, there is another simple solution we can easily
guess, based on the properties of the Riemann tensor.
If we assume that nothing depends on the extra dimension, 
we can consider a spacetime of the form
\be
ds^2 = g_{\mu\nu} (x^\mu) dx^\mu dx^\nu + dz^2\;,
\ee
for which the Riemann tensor has only four dimensional
components. In this case, a solution
of the four-dimensional Einstein equations will
automatically be a solution of the five-dimensional
Einstein equations, as $R_{5a}=0$ by construction.
Thus, we can extend the four-dimensional
Schwarzschild solution uniformly into the
extra dimension to obtain a black string: \index{black string|(}
\be
ds^2 = - V(r) dt^2 + V(r) ^{-1} dr^2 + r^2 d\Omega_2^2 + dz^2\;,
\label{bsmetric}
\ee
where 
\be
V(r) = \left ( 1 - \frac{r_+}{r} \right )
\ee
is the Schwarzschild potential introduced in the previous
chapter.

This may seem a rather unremarkable observation, as it
is a straightforward solution to write down. However, it
is the first sign that gravity in higher dimensions may have
distinctively new phenomena to offer. In four dimensions,
black holes are essentially unique, classified by very
few parameters: mass, charge and angular momentum.
Once these parameters are specified, the solution is
known, and the horizon topology is always spherical.
Here, by adding an extra dimension, with very little effort we
have constructed two distinct black objects: one with a
spherical and one with a cylindrical event horizon. Clearly,
these black holes have different masses: the black string 
has (strictly) infinite mass and is not asymptotically flat,
however, our simple construction shows that event horizons in 
higher dimensions need no longer be spherical \cite{topology2}; 
Chapter 7 discusses the possible topologies of black
objects. As we will see in Chapters 6 and 8, this first clue
of non-uniqueness of `black' solutions is the tip of the
iceburg: \index{non-uniqueness} many distinct black solutions with the same charges exist,
see e.g.\ \cite{ring2,black2}.

\begin{figure}[ht]
\includegraphics[width=0.7\textwidth]{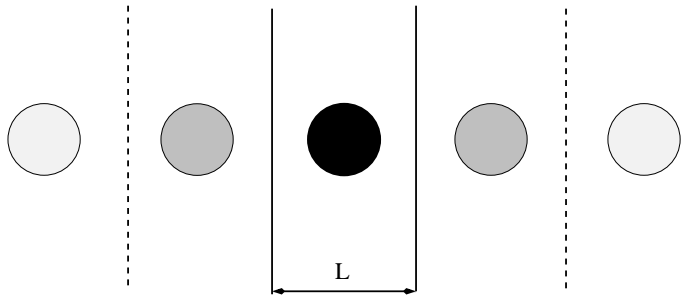}
\caption{A sketch of the five dimensional black hole
confined within a finite extra dimension, and its
mirror images with which it `interacts'.}
\label{fig:caged}
\end{figure}

Now let us consider what happens if our extra dimension 
is compact, i.e., finite and of length $L$. This will give the
black string a finite length, hence mass, and in fact
corresponds to the traditional Kaluza-Klein picture. The black
string therefore corresponds to a basic Kaluza-Klein black hole;
there is no dependence of the geometry on the extra
dimension, which remains a Killing symmetry of the
full solution. From the four dimensional viewpoint, it looks just like a Schwarzschild black hole.
At energies of order $L^{-1}$,
new physics is expected to come into play $-$ physics
corresponding to the additional degrees of freedom
of the extra dimension. In our case, as we are considering
only vacuum Einstein gravity, our new degrees of
freedom correspond to a dependence of the geometry 
on the extra dimension. Alternatively, from a four
dimensional perspective, these can be interpreted 
as a tower of massive gravitons. We can therefore 
ask if there are any alternative solutions to the black
string which excite these extra degrees of freedom.
An obvious starting point is the five dimensional
`Schwarzschild' solution, (\ref{5DSCH}). Once we have a 
finite spatial direction, we do not expect exact hyperspherical 
symmetry, since the finite size of the extra dimension
introduces an effective periodicity in one direction,
and hence the black hole will interact with its mirror image 
black holes (see Fig. \ref{fig:caged}), altering the gravitational 
potential along the extra dimension. 
For $r_5 \ll L$, the
five dimensional potential (\ref{5Dpot}) will be a good approximation
to the solution, but for larger black holes, the nonlinearity of 
gravity does not allow for an analytic solution, and the
geometry must be found numerically. As the mass of the black
hole increases further, it eventually can no longer fit inside the
extra dimension, and must become a string-like solution.
These solutions are known as {\it caged black holes}, and
{\it nonuniform black strings}, and will be discussed further in
Chapter 4 (see also \cite{caged2}).

Therefore, within the context of five dimensional Kaluza-Klein theory, there seem to be various options for a simple
uncharged black hole. It can be a straight or nonuniform
black string, or a caged black hole, but which is the
physically relevant one, or which will form in a collapse
process? Are all of these solutions possible, or is there a
selection process which rules out one or the other?

The answer is provided by the black string instability: 
dependent on the size of the extra dimension
and the mass of the black hole, there is a unique stable solution, 
and hence a unique preferred end state for gravitational collapse.
The rest of this chapter is devoted to explaining why the instability
is natural, and proving that it exists.

\section{A thermodynamic argument for instability}\index{instability!black string|(}
\index{black hole thermodynamics}

Typically in nature, we decide which is the most likely state by
determining which has the lowest energy, however, in the
case of black holes and black strings, both solutions can have
the same energy, therefore a different physical principle must
be used. In the last chapter, we saw how black holes could be
assigned thermodynamic properties, and in thermodynamics 
it is the state with largest entropy which is preferred. Thus, if
there is an entropy difference between the two states, then
we might expect that one is preferred over the other.

In the previous chapter, the entropy of a black hole was shown to 
be proportional to its area: $S = A/4$ in Planck units. 
However, there is a subtlety if we are dealing with Kaluza-Klein theory; this
formula actually contains a hidden Newton's constant, and the
Planck mass is dimension dependent in a compactification:
\be
M_{p}^2 = V_{D-4} M_D^{D-2}\;,
\ee
where $V_{D-4}$ is the volume of the internal space on which we are
compactifying, and $D$ is the total number of dimensions. For the five 
dimensional case we are considering here, this gives
\be
G_5 = L G_4 = L\;,
\label{planckmass}
\ee
(where we have set $G_4 =1$ in the last step) and hence what we mean by the Planck scale is renormalized by a factor
of $L$.  We therefore obtain for the entropies of the black hole
and black string (assuming, for simplicity, that the black hole 
is approximated by its exact analytic form (\ref{5DSCH})): \index{black hole thermodynamics!entropy}
\be
S_{BH} = \pi^2 r_5^3 / 2L \;\;\;\;,\;\;\;\;\;
S_{BS} = \pi r_+^2 \;,
\ee
where $r_5$ and $r_+$ are the horizon radii of the black hole and string
respectively. We now need to compare these entropies for the same total
mass of hole or string. The black string has a mass of $r_+/2$, 
and using (\ref{M5}) and (\ref{planckmass}), the black hole mass is
\be
M_5 = \frac{3\pi r_5^2}{8L}\;.
\ee
Hence, setting the masses equal,  the entropies can be re-expressed as:
\be
S_{BH} = 4\pi M^2 \sqrt{\frac{8L}{27\pi M}}\;\;\;\;,\;\;\;\;\;
S_{BS} = 4\pi M^2\;\;\;\;.
\ee
Clearly, if $L$ becomes sufficiently large, the black hole
will be thermodynamically preferred over the black string, and hence
the black string solution should have a long wavelength instability.
In order to prove this we have to look at perturbations
around the black string solution.

\section{Perturbing the black string}

Now we will show explicitly that the black string is unstable. In order to
do this, we perturb the metric (\ref{bsmetric}), and solve the linearized
Einstein equations to show that there is a growing mode. 

There are several issues to bear in mind when considering perturbation
theory in general relativity. First of all, a perturbation must be ``small'', this may seem to
be a statement of the obvious, but when a change of coordinates
can make the components of a tensor large, we must be careful
to interpret correctly what ``small''  means. Secondly, 
we need to specify an initial data surface for our perturbation problem,
which as we will see ties in with regularity of the perturbation, and is 
easily resolved by choosing an appropriate Cauchy surface.
Finally, gravity has an infinite gauge group, i.e., there are
an infinite set of different coordinate transformations we can
perform on any particular geometry, thus there will be many perturbations
which are pure gauge -- in other words, the act of  changing coordinates gives
a perturbation to the metric, but one that is not physical. We must therefore
be careful to determine whether our perturbation is physical. 

\subsection{Perturbation theory}

We begin by defining the perturbation. In Einstein gravity, a small
perturbation of a spacetime is represented by a change in the metric:
\be
g_{ab} \to g_{ab} + h_{ab}
\ee
under which the Ricci tensor acquires a perturbation
\be
R_{ab} \to R_{ab} - \frac12 \Delta_L h_{ab}\;,
\ee
where $\Delta_L$ is the Lichnerowicz operator, \index{Lichnerowicz operator}
\be
\Delta_L h_{ab} = \Box h_{ab} + 2 R_{acbd} h^{cd} - 2 R^c_{(a} h_{b)c}
- 2\nabla_{(a} \nabla^c h_{b)c} + \nabla_a \nabla_b h\;,
\label{Lichop}
\ee
the curved space wave operator for a spin two massless particle.
Clearly, a perturbation of a vacuum spacetime must obey $\Delta_L h_{ab}=0$.

For the black string, the facts that the Ricci tensor is zero (as the string 
is a solution of the vacuum Einstein equations), and that there are 
no $z$ components of the Riemann tensor, will simplify the equations
considerably. In addition, since we are in vacuum we can also 
choose the ``transverse tracefree'' gauge for $h_{ab}$\footnote{If
not in vacuum, we can still choose $\nabla_ah^a_b - \frac12\nabla_bh=0$,
just not the individual parts separately.}, \index{transverse tracefree gauge}
\be
\nabla_a h^a_b = 0 = h \;,
\label{TTF}
\ee
which further simplifies (\ref{Lichop}) to
\be
\Delta_L h_{ab} \to \Box h_{ab} + 2 R_{acbd} h^{cd} = 0\;.
\label{Lichred}
\ee
It is now a matter of some algebraic computation and manipulation
to compute the perturbation equations component by component, using
the gauge choice to simplify equations where relevant.

As is standard practice, we use a separation of variables method,
and decompose the perturbation in terms of the symmetries, or
Killing vectors, of the background geometry. The black string has 
both time and $z-$translation invariances, as well as an $SO(3)$
isometry corresponding to the four dimensional spherical
symmetry of the Schwarzschild solution. For simplicity (and
with the benefit of hindsight!) we consider spherically symmetric 
perturbations, since the entropy argument indicates that the instability
should manifest at this level. This means that $h_{ab}$ has no cross
terms with an angular coordinate, and has the form:
\be
h_{ab} = 
\left [
\begin{matrix}
h_{tt} & h_{tr} & 0 & 0 & h_{tz} \\
h_{tr} & h_{rr} & 0 & 0 & h_{rz} \\
0 & 0 & h_{\theta\theta} & 0 & 0 \\
0 & 0 & 0 & h_{\theta\theta} \sin^2\theta & 0 \\
h_{tz} & h_{rz} & 0 & 0 & h_{zz} 
\end{matrix}
\right ]\;\;\;.
\ee
In addition, the $t-$ and $z-$translation symmetries allow us 
to factor out an oscillatory $e^{im  z}$ behaviour, and a
growing $e^{\Omega t}$ mode, corresponding to an
unstable perturbation.

\subsection{Finding the perturbation}

The full set of equations is rather lengthy, and not particularly 
illuminating, so we refer the reader to the original literature
for the details, \cite{Gregory2:1993vy}. For the purposes of this review, we note
that the perturbations $h_{zz}$ and $h_{z\mu}$ must
vanish for any unstable mode. To see this is particularly
straightforward for the $h_{zz}$ perturbation as the lack of
Riemann components in the $z$-direction means that the
equation for this component decouples. 
Writing $h_{zz} = e^{im  z} e^{\Omega t} h$ we obtain:
\be
h'' + \left (\frac{2r-r_+}{r-r_+} \right ) \frac{h'}{r}
- \left ( m ^2 r(r-r_+) + \Omega^2 r^2 \right ) \frac{h}{(r-r_+)^2} = 0\;.
\label{hzzeq}
\ee
This equation has asymptotic solutions
\bea
h &\sim& e^{\pm \sqrt{\Omega^2+m ^2}\,r} \qquad {\rm as} \qquad
r\to\infty \;,\nonumber \\
h &\sim& (r-r_+)^{\pm \Omega r_+}\qquad {\rm as} \qquad
r\to r_+ \;. \nonumber 
\eea
Clearly therefore, any regular solution must vanish at the horizon
and infinity, with a turning point at some finite $r$ at which $h''/h<0$. 
However, examination of (\ref{hzzeq}) shows that $h''/h>0$ at any
turning point, hence no such solution exists.
A similar argument shows that $h_{z\mu}$ must also vanish.

We are now left with a perturbation which has only four dimensional
components, $h_{\mu\nu}  = e^{im  z} e^{\Omega t} H_{\mu\nu}(r)$. 
After imposing the gauge constraints, the equations of motion 
reduce to a pair of first order ODE's plus one constraint:
\bea
H_+  & = &
\frac{H_-}{V} \frac{\left(2 r^2 \Omega^2 + r^2m ^2 V - (1-V^2)/2 \right )}
{\left (r^2 m ^2 +1-V \right ) }
-\frac{rH}{\Omega} \frac{\left ( 4\Omega^2 + m ^2 (1-3V) \right )}
{\left (r^2 m ^2 +1-V \right ) } \qquad\\
H' &=& \frac{\Omega(H_+ + H_-)}{2V} - \frac{(1+V) H}{rV} 
\label{Heq}\\
H_-' &=& \frac{m^2 H}{\Omega} + \frac{H_+}{r}
+\frac{(1-5V)H_-}{2rV} \label{Hmeq}
\eea
in the variables
\bea
H_{\pm}  &=&  \frac{H_{tt}}{V}  \pm V H_{rr} \\
H &=& H_{tr}\;.
\eea
Once again, reading off the asymptotic behaviour gives
\bea
r\to\infty \;\;\; :\;\;\; &&
\begin{cases}
H \;\sim \pm \sqrt{\Omega^2+m^2} \; e^{\pm \sqrt{\Omega^2+m^2}\,r}  \\
H_- \sim \frac{m^2}{\Omega} e^{\pm \sqrt{\Omega^2+m^2}\,r} 
\end{cases}
\qquad \qquad \label{infinity} \\
r\to r_+  \;\;\; :\;\;\; &&
\begin{cases}
H\; \sim \left (\pm \Omega r_+ - \frac{1}{2} \right ) (r-r_+)^{\pm \Omega r_+-1}\\
H_- \sim \left (\frac{m^2}{\Omega} \pm \frac{2}{\,r_+} \right ) 
(r-r_+)^{\pm \Omega r_+}\;.
\end{cases}
\label{horizon} 
\eea
An instability therefore corresponds to a solution of 
(\ref{Heq}), (\ref{Hmeq}) which is regular at both the
horizon and infinity, as determined by the asymptotic forms
(\ref{infinity}), (\ref{horizon}).

\subsection{Regularity conditions}

\begin{figure}[ht]
\includegraphics[width=0.7\textwidth]{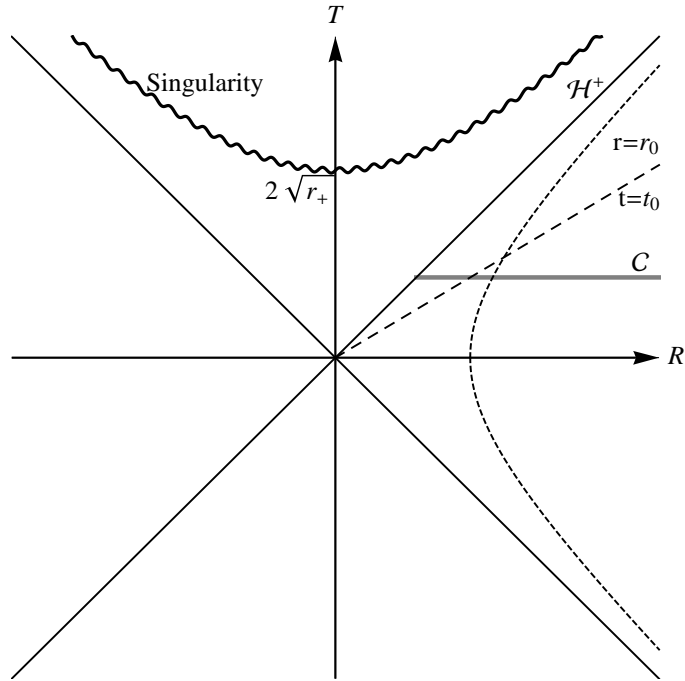}
\caption{A diagram of the black hole spacetime in 
Kruskal-style coordinates $(R,T)$. The future event horizon,
${\cal H}^+$ and singularity are labelled, as well as lines of
constant $t$ and $r$, the original Schwarzschild
coordinates. The initial data surface, ${\cal C}$, from which we 
evolve the perturbation is also schematically indicated.}
\label{fig:ids}
\end{figure}

In order to determine the regularity of the perturbation, 
we clearly need $h_{ab}$ to tend to zero at large $r$,
picking out the exponentially decaying branch of (\ref{infinity}),
and also to be regular at the horizon. Surprisingly, this latter 
constraint is not equivalent to the perturbation being regular in
a local orthonormal system, as it is easy to see from
(\ref{horizon}) that $h_{{\hat t}{\hat r}} = h_{tr}$ blows up
as $r\to r_+$ for $\Omega r_+ <1$. Instead, we have to check 
regularity in a locally regular coordinate system at the 
horizon. A convenient choice of coordinates is
based on the Kruskal system: \index{Kruskal coordinates|(}
\be
T = 2 e^{r_*/2r_+} \sinh ( \frac{t}{2r_+} ) \;\;\;\;,\;\;\;\;\;
R = 2 e^{r_*/2r_+} \cosh ( \frac{t}{2r_+} )\;\;\;,
\ee
where
\be
r_* = r - r_+  +  r_+ \log (r-r_+)
\ee
is the standard tortoise coordinate in the Schwarzschild metric
(see Fig.  \ref{fig:ids}).

Transforming to this new coordinate system, we see that
\bea
h_{TT} &\sim& \frac{{\cal U}(R,T)}{(R^2-T^2)} \left (
R^2 h_{{\hat t}{\hat t}} - 2RT h_{{\hat t}{\hat r}}
+T^2 h_{{\hat r}{\hat r}}  \right) \\
h_{TR} &\sim& \frac{{\cal U}(R,T)}{(R^2-T^2)}  \left (
RT (h_{{\hat t}{\hat t}} +h_{{\hat r}{\hat r}})
- (T^2+R^2) h_{{\hat t}{\hat r}} \right) \\
h_{RR} &\sim& \frac{{\cal U}(R,T)}{(R^2-T^2)}  \left (
T^2 h_{{\hat t}{\hat t}} - 2RT h_{{\hat t}{\hat r}}
+R^2 h_{{\hat r}{\hat r}} \right) 
\eea
must be regular at the horizon, where $h_{{\hat t}{\hat t}}=h_{tt}/V$ 
refers to the components in a local orthonormal system, and
\be
{\cal U}(R,T) = {\cal U} \left (R^2-T^2\right ) 
= \frac{r_+^2}{re^{(r-r_+)/r_+}}
\ee
is the new (non-singular) Kruskal gravitational potential.

Substituting in the near horizon behaviour gives
\bea
h_{TT} \simeq h_{RR} &\propto& \pm \left (R\pm T\right )^{\Omega r_+ -2}\\
h_{TR} &\propto&  \left (R\pm T\right )^{\Omega r_+ -2}
\eea
and we see that the upper branch of (\ref{horizon}) is 
always regular on the future event horizon ($R=T$), and the 
lower branch on the past event horizon. At the bifurcation
point, where ${\cal H^+}$ and ${\cal H}^-$ meet,
corresponding to $r=r_+$ at finite $t$, neither branch
is strictly regular, and to exclude both would render the
Lichnerowicz operator non self-adjoint.
For simplicity, it is easiest simply to consider perturbations that are
regular on the future event horizon, as the black string is
presumed to form from gravitational collapse, and hence 
any initial data surface would have to be chosen to 
terminate on the future horizon. Fig. \ref{fig:ids}
shows the black hole spacetime in Kruskal coordinates,
together with the initial data surface.
\index{Kruskal coordinates|)}

\subsection{The instability}

\begin{figure}[ht]
\includegraphics[width=0.7\textwidth]{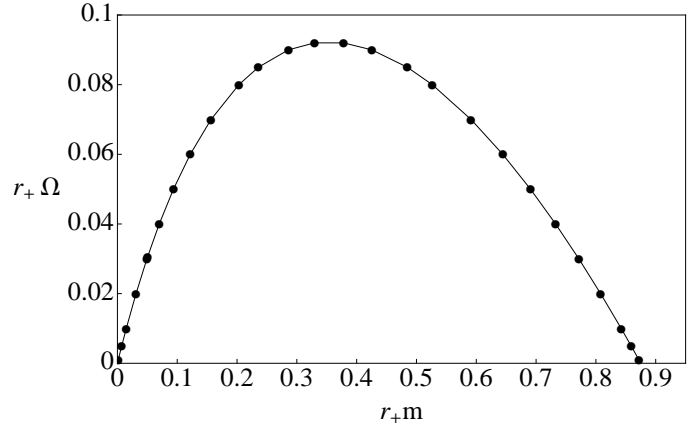}
\caption{A plot of the eigenvalues $(m,\Omega)$,
scaled by $r_+$, for which an instability is present. }
\label{fig:evals}
\end{figure}

\begin{figure}[ht]
\includegraphics[width=0.7\textwidth]{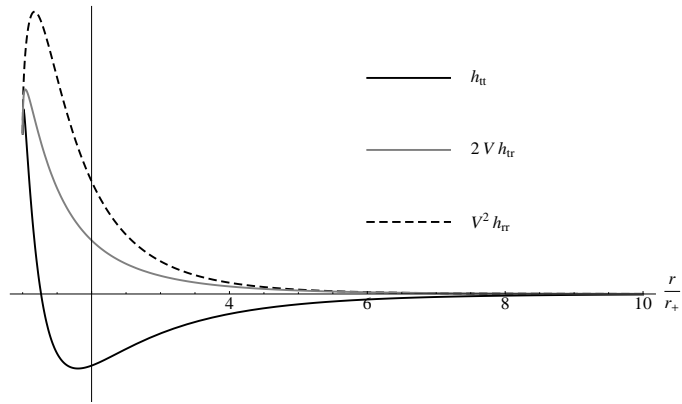}
\caption{A plot of the metric perturbation.}
\label{fig:efns}
\end{figure}

To determine the existence of the instability we
must numerically integrate the perturbation equations
(\ref{Heq}) and (\ref{Hmeq}) between the horizon and infinity, 
looking for a solution which approaches the regular horizon
branch, and is exponentially decaying at infinity. 
We do not expect solutions for all $\Omega$ and $m$, 
since the thermodynamic argument indicates that an 
instability can only set in for $r_+m < 32/27$. We
expect a single characteristic 
frequency, $\Omega_m$, for any wavelength, thus
we must scan through the values of $\Omega$ for
each $m$ to check if a solution exists.
Fig. \ref{fig:evals} shows a plot of the frequency
pairs $(m,\Omega)$ for which a regular solution,
and hence an instability, exists, and Fig. \ref{fig:efns}
shows the behaviour of the perturbation.

Having found an unstable solution to
the perturbation equations, the final step of the argument
is to demonstrate that this is a physical instability of
the black string, and not just some odd gauge mode. In fact,
this is easy to demonstrate by looking at (\ref{Lichred}).
Since both the perturbation and the Riemann tensor vanish
in the extra dimension ($h_{za} = 0 = R_{zabc}$), the five
dimensional Lichnerowicz operator reduces to the four
dimensional Lichnerowicz operator with a mass term:
\be
\Delta^{(5)}_L h_{\mu\nu} = \Delta_L^{(4)} h_{\mu\nu} + 
\frac{\partial^2 \;}{\partial z^2} h_{\mu\nu} = \left [
\Delta_L^{(4)} - m^2 \right ] h_{\mu\nu}\;.
\label{massiveLich}
\ee
However, if $h_{\mu\nu}$ is a gauge mode, it must correspond
to a purely four dimensional change of coordinates, in other
words, it can have {\it no} dependence on $z$. Thus any solution
of the {\it massive} four dimensional Lichnerowicz operator must
be a physical Kaluza-Klein instability.

\subsection{More general instabilities}

So far, the discussion has been strictly in terms of
five dimensional vacuum Einstein gravity. This approach
was chosen so that the mathematics and physics of the
instability would be clearer, but of course if an instability
exists with one extra dimension, then it will exist more
generally. In \cite{Gregory2:1993vy}, it was shown how black branes, 
\index{black brane}
objects with arbitrary numbers of extra dimensions, would
be unstable, with $1-6$ extra dimensions focussed on for
the purposes of applying to the string theoretic solutions
found by Horowitz and Strominger \cite{HS2}. \index{instability!black brane}

These instabilities look very similar to
the five dimensional case detailed here: the instability
is once more restricted to a four dimensional s-wave, where
now the effective mass term in (\ref{massiveLich}) is a 
general eigenvalue of the symmetries in the extra dimensions,
$e^{m_iz^i}$. The details of the $(m,\Omega)$ plot vary, but
the qualitative shape and features are the same (see \cite{Gregory2:1993vy}).
Instabilities of charged solutions, analogous to the 
four dimensional Reissner Nordstrom family of black holes
(see Chapter 11) can also be found.

Interestingly, the instability does not require a translation
invariance along the string or brane, it also applies to more
general higher dimensional spacetimes. Essentially, all that 
is required is some sort of factorizability of the metric and wave
operator \cite{RGL2}, so that we can decompose
the perturbation in terms of effective four dimensional
quantities with suitable eigenfunctions of the extra 
dimensions:
\be
h_{ab} \to h_{\mu\nu} = u_m(z^i) e^{\Omega t} H_{\mu\nu}(r)
\ee
where the Riemann tensor and wave operator also factorize
so that a massive wave equation in the form of (\ref{massiveLich})
is obtained for $H$. 

\section{Consequences of the instability}

In the previous section, we proved that an instability of the black
string exists, in that there is a linear perturbation of the 
black string solution which is exponentially growing in our
coordinate time, $t$. However, it is not clear what the effect of
this growing mode will be on the event horizon, which is a
coordinate singularity, and in fact corresponds to $t\to\infty$,
$r\to r_+$ in Schwarzschild coordinates.

\begin{figure}[ht]
\includegraphics[width=0.7\textwidth]{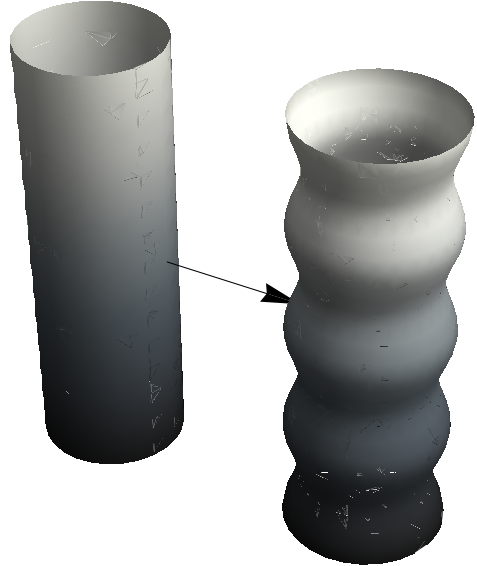}
\caption{A representation of the effect of the instability
on the black string horizon.}
\label{fig:evol}
\end{figure}

To explore the effect of the instability, we return to the 
Kruskal coordinates and check what happens to outgoing light
rays near the original event horizon. In the unperturbed
spacetime, null geodesics satisfy $R = \pm T +R_0$, with
$R=T$ being the future event horizon, as indicated in Fig.
\ref{fig:ids}. In the perturbed spacetime, the geodesic
equation becomes
\bea
\left ( \frac{dR}{dT} \right )^2 &=& 1 +
\frac{1}{{\cal U}} \left ( h_{TT} + 2 h_{TR} {\dot R} 
+ h_{RR} {\dot R}^2 \right ) \\
&=& 1 + \epsilon \cos m z (R+T)^{2r_+\Omega -2} 
\left ( 1 + \frac{dR}{dT} \right )^2
\eea
where $\epsilon$ is an (arbitrary) small parameter
representing the size of the initial perturbation.
From this, we see that the event horizon is schematically
shifted to
\be
R = T + \epsilon \cos m z T^{2r_+\Omega-1}
\ee
or, in Schwarzschild coordinates
\be
r = r_+ + \epsilon T^{2\Omega} \cos m z
\ee
in other words, the `horizon' begins to ripple (see Fig. \ref{fig:evol}).

This now has very interesting consequences. In four dimensional
gravity, the event horizon cannot shrink in any classical
process without violating positivity of energy. This black
string instability is a classical process, so what is happening?
Clearly, although the horizon is shrinking in some places along the black
string, in other places it is growing, hence overall this classical
process is growing the total area, as we would expect from
the thermodynamic argument of entropic instability. However, increasing
the area of an event horizon is not the only classical relativistic
constraint. If we follow the instability to its logical endpoint,
dictated by the entropy argument, we might expect that the black string
will eventually fragment, forming a black hole caged within the 
larger fifth dimension. A simple conclusion perhaps, but at the moment 
when the horizon pinches off, the curvature at the horizon diverges, 
forming a naked singularity.

In the catalogue of four dimensional relativity, 
a pattern emerges where, apart from a few very well known 
examples such as the Big Bang, singularities tend to be 
clothed by an event horizon, and certainly singularities which
form during a physical collapse. This led Penrose to conjecture 
\cite{CC2} that a `censorship' applies in gravitational
collapse, which prevents any singularity forming which could
be visible from  infinity. Although a full
proof of this conjecture remains elusive, any counter-examples that
have been constructed are either unphysical in some way, or highly
non-generic. The {\it Cosmic Censor} therefore has been assumed to
be an omnipotent authority in classical gravity. \index{cosmic censorship} 
Yet here, within
the bounds of classical gravity, a physical, generic process has
been shown to exist which strongly suggests a violation of cosmic
censorship at the moment the string fragments into the black hole.
For this reason, for many years after the discovery of the black
string instability, the final fate of the black string was viewed as
 an open question \cite{fate2}, and cosmic censorship was an unknown
factor in higher dimensional gravity. The story of what happens to
the black string, and how the instability proceeds requires a 
tour de force numerical simulation \cite{finalstate2}, a
description of which forms the core of the next chapter.
\index{black string|)}
\index{instability!black string|)}

\end{document}